# EPS Confidentiality and Integrity mechanisms Algorithmic Approach

**Ghizlane ORHANOU, Saïd EL HAJJI, Youssef BENTALEB and Jalal LAASSIRI**

**Département Mathématiques et Informatique, Laboratoire Mathématiques Informatique et Applications, Université Mohammed V - Agdal, Faculté des Sciences, BP 1014, Rabat, Maroc**

**Abstract**

The Long Term Evolution of UMTS is one of the latest steps in an advancing series of mobile telecommunications systems. Many articles have already been published on the LTE subject but these publications have viewed the subject from particular perspectives. In the present paper, a different approach has been taken. We are interested in the security features and the cryptographic algorithms used to ensure confidentiality and integrity of the transmitted data. A closer look is taken to the two EPS confidentiality and integrity algorithms based on the block cipher algorithm AES: the confidentiality algorithm EEA2 and the integrity algorithm EIA2. Furthermore, we focused on the implementation of both algorithms in C language in respect to the specifications requirements. We have tested our implementations according to the testsets given by the 3rd Generation Partnership Project (3GPP) implementation document. Some examples of the implementation tests are presented bellow.

## 1. EPS Security Mechanisms

EPS (*Evolved Packet System*) represents the very latest evolution of the UMTS standard. EPS is also known by other acronyms related to technical study items being worked on at 3GPP committees: **LTE** (*Long Term Evolution*), which is dedicated to the evolution of the radio interface, and **SAE** (*Service Architecture Evolution*) which focuses on Core Network architecture evolution.

EPS is specified as part of the 3GPP family and proposes a significant improvement step, with a new radio interface and an evolved architecture for both the Access and the Core Network parts.

Security is another important feature of the 3GPP family and its evolution is an important issue. EPS provides security features in a similar way as its predecessors UMTS (Universal Mobile Telecommunication System) and GSM (Global System Mobile). In addition to the mutual authentication functionality of the network and the user, two other security functions are provided to ensure data security during its transmission over the air interface and through the LTE-SAE system: ciphering of both user plane data and control plane data (in the RRC (Radio Resource Control) layer), and integrity protection which is used for control plane data only. For the NAS (Non-Access Stratum) network, both ciphering and integrity are provided.

Ciphering is used in order to protect the data streams from being received by a third party, while integrity protection allows the receiver to detect packet insertion or replacement.

In the present paper, we will focus on the study of EPS confidentiality and integrity mechanisms and on the cryptographic algorithms used to fulfill these security features.

### 1.1 LTE Confidentiality and Integrity Layer

User and signaling data are considered sensitive and their confidentiality and integrity should be protected between the UE (*User Equipment*) and the serving network. For this reason and in contrast with UMTS (*Universal Mobile Telecommunication System*) where the data confidentiality and integrity had been ensured only in the air interface (between UE and RNC (*Radio Network Controller*)), these features in the EPS have been implemented in different levels to ensure more data security.

### 1.1.1 Overview of LTE control and user plane protocol stacks

Confidentiality and Integrity protection for RRC (*Radio Resource Control*) and UP (*User Plane*) data is provided between the UE and the e-NB (*Evolved Node-B*) in the Access Stratum (AS). These security features are applied at the PDCP (*Packet Data Convergence Protocol*) layer, and no layers below PDCP are confidentiality protected. For the NAS signaling protection, the security controls are done between UEs and MMEs (*Mobility Management Entity*) by the NAS protocol [1, 2].

The PDCP layer is the upper sub-layer of Layer 2 in the LTE Protocol stack. Fig.1 bellow shows the user plane protocol stack in the Enhanced UMTS Terrestrial Radio Access Network (E-UTRAN) named also LTE.

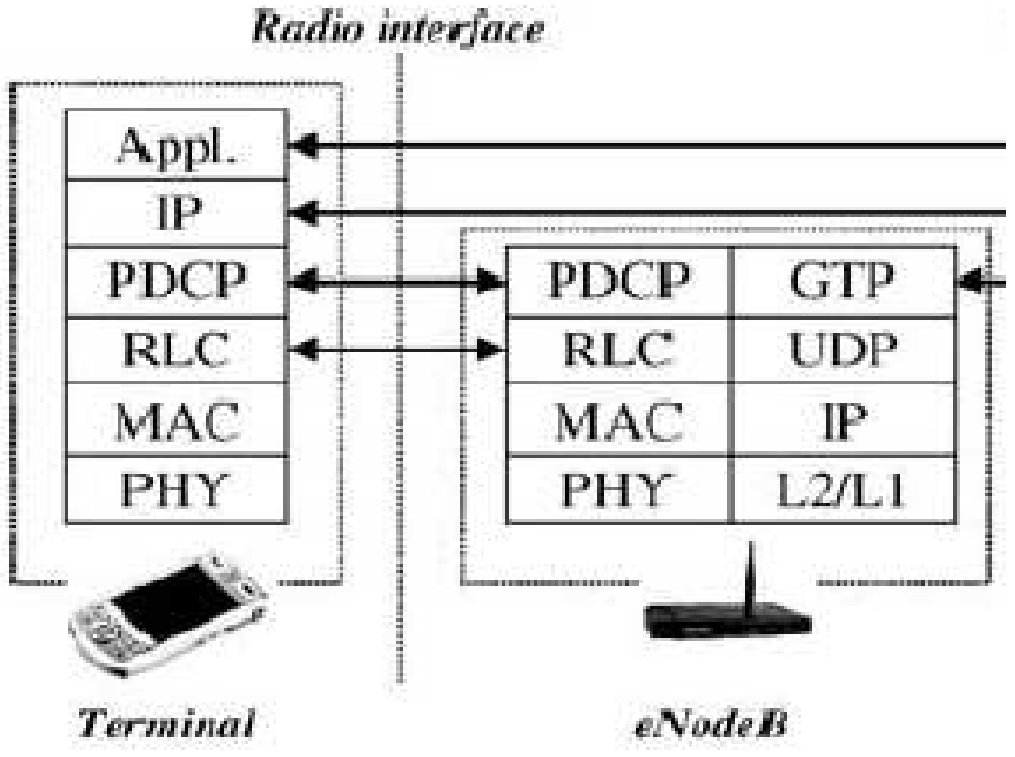

Fig. 1 LTE User plane protocol stack [3]

On the other hand, Fig. 2 bellow shows the emplacement of the RRC and NAS signaling in the EPS control plane.

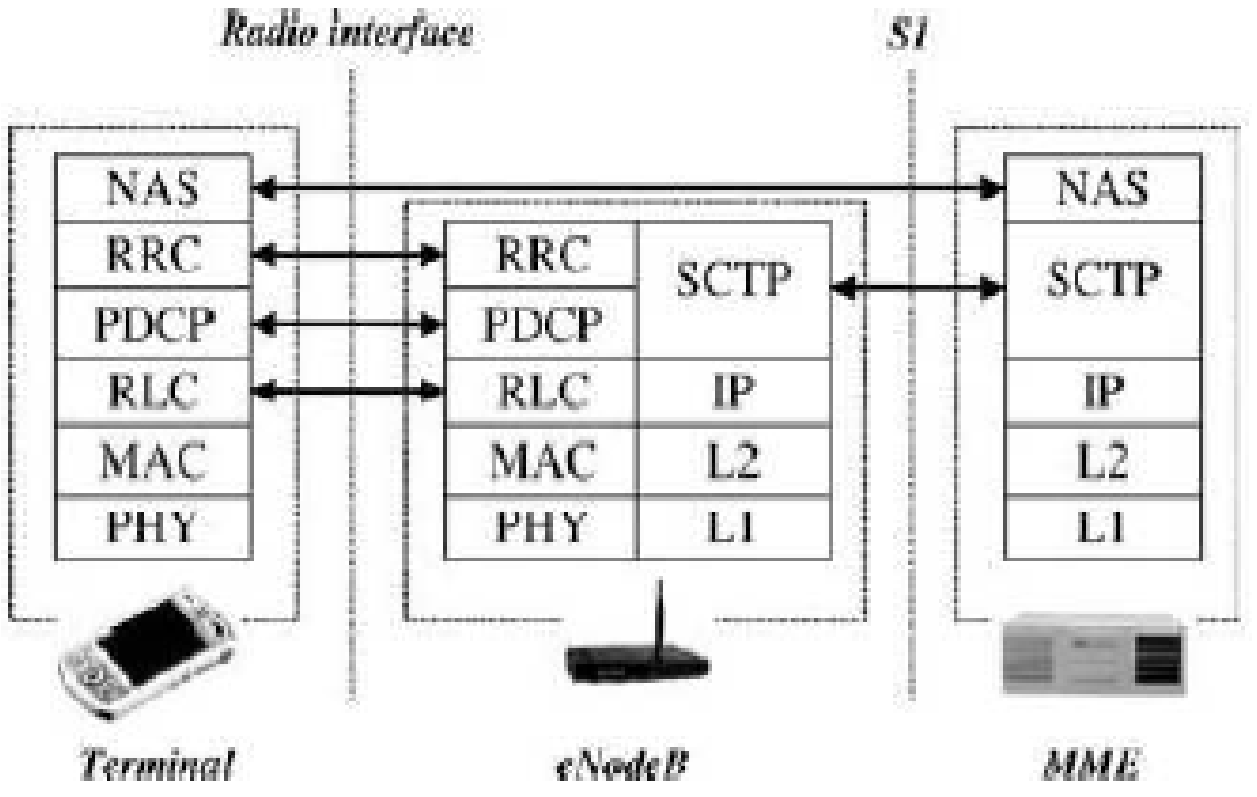

Fig. 2 EPS Control plane protocol stack [3]

After presenting control and user plane stack, we will present in the following subsection an overview of the implementation of integrity and confidentiality features in the PDCP layer.

### 1.1.2 PDCP security functions

The PDCP layer manages data streams for the user plane, as well as for the control plane. The architecture of the PDCP layer differs for user plane data and control plane data, as shown in the figures Fig. 3 and Fig. 4 respectively.

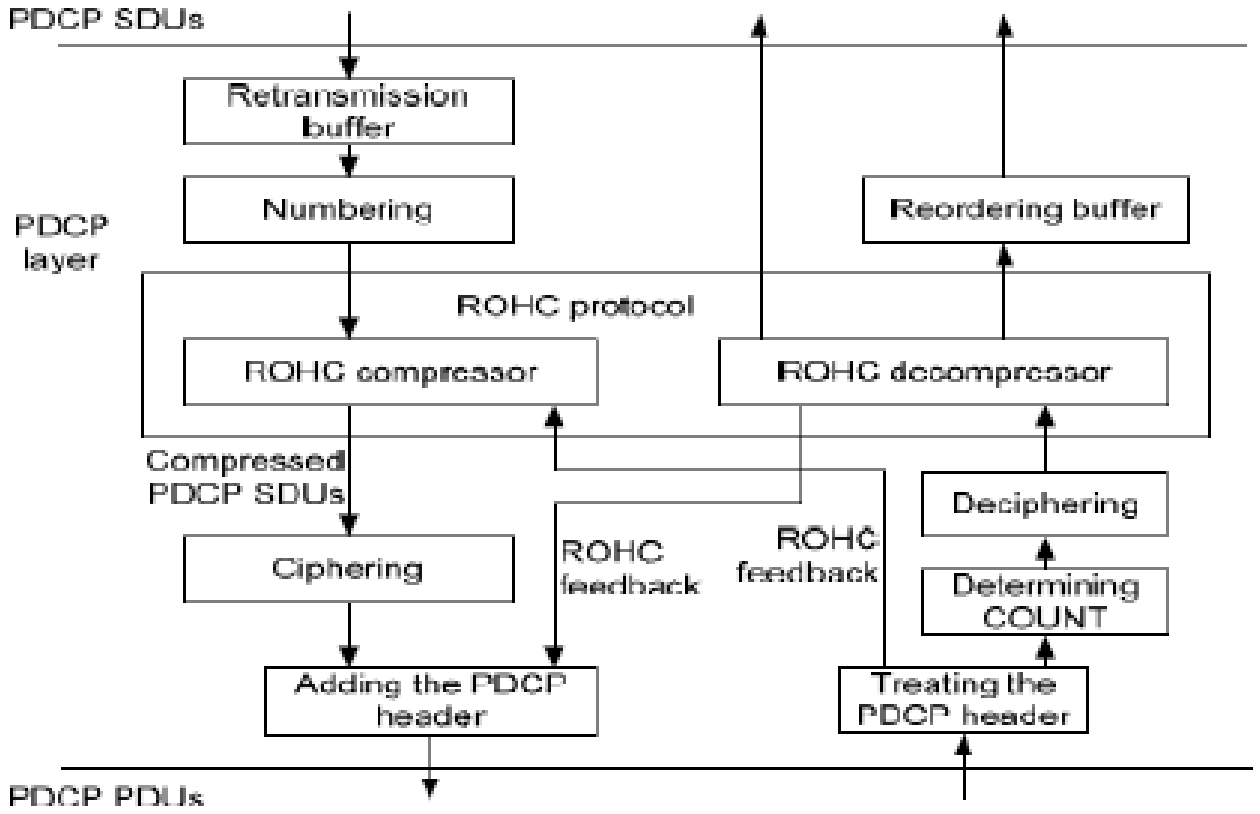

Fig. 3 Overview of user-plane PDCP [1, 4]

The ciphering function performed by the PDCP layer includes both ciphering and deciphering. For the user plane, the data unit that is ciphered is the data part of the PDCP PDU, as shown in the figure Fig. 3 above.

For the control plane, the data unit that is ciphered is the data part of the PDCP PDU and MAC-I. Indeed, PDCP Data PDUs for control plane data comprise a MAC-I field of 32-bit length for integrity protection as presented in Fig. 4 bellow.

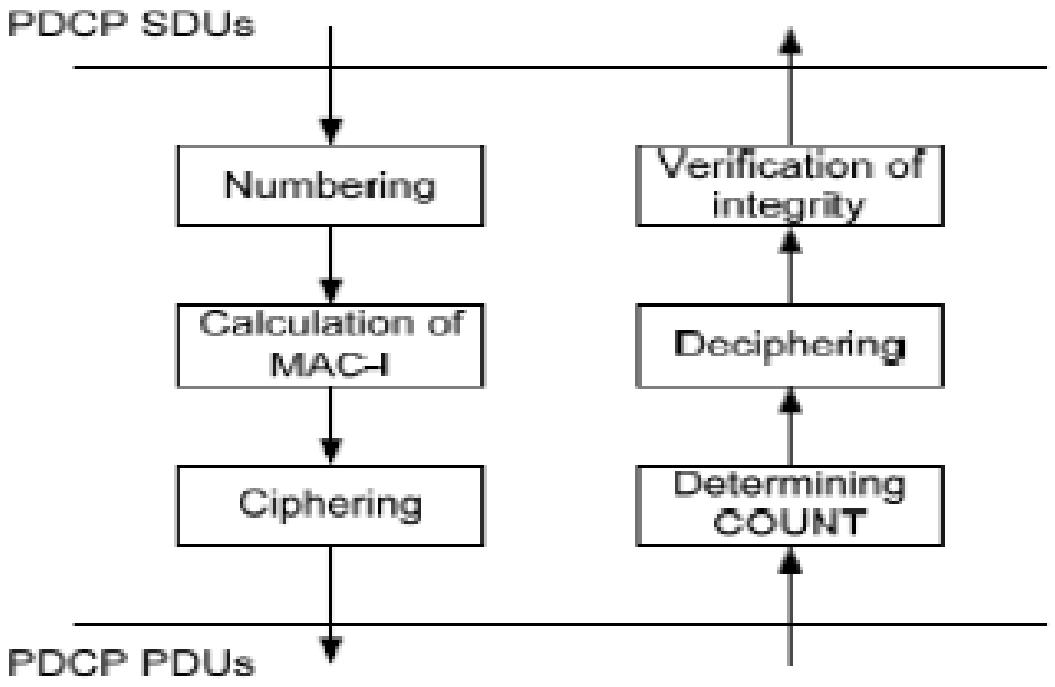

Fig. 4 Overview of control-plane PDCP [1, 4]

Moreover, it is important to mention that the NAS, independently, applies integrity protection and ciphering in the NAS layer.

The ciphering and integrity algorithms and keys to be used by the PDCP entity are configured by upper layers.

## 1.2 Confidentiality and Integrity mechanisms

### 1.2.1 User and signaling data Confidentiality

To ensure the data confidentiality, the following procedures are provided [5]:

- Cipher algorithm EEA (*EPS Encryption Algorithm*) agreement: To ensure the confidentiality of user and signaling data in LTE-SAE (*Long Term Evolution - Service Architecture Evolution*), 3GPP has maintained the use of the UMTS algorithm UEA2 based on SNOW 3G algorithm [1, 6], and has named it EEA1. In addition, a new algorithm EEA2, based on AES algorithm used in the CTR mode (*Counter Mode*), has been adopted. We will be interested in its operation study in detail in section 2. of the present paper. Besides, the UE and the EPS can securely negotiate the algorithm to use in their mutual communication.

- Cipher key agreement: the agreement is done between the UE and the network during the Authentication and Key Agreement procedure;

- Encryption/Decryption of user and signaling data;

### 1.2.2 Signaling data Integrity

Data integrity in the EPS network ensures the protection of the signaling data integrity and allows the authentication of the signaling messages transmitted between the user and the serving network [5, 7]. User data is not integrity protected.

Integrity protection, and replay protection, shall be provided to all NAS and RRC-signaling messages except those explicitly listed in 3GPP documents [1].

The following security features are provided to ensure the signaling data integrity on the LTE and SAE:

- Integrity algorithm (EIA) agreement: as for the data confidentiality, there are actually two variants of the integrity algorithm for LTE: EIA1 based on SNOW 3G algorithm (named UIA2 in UMTS network) [1, 6]. It was used since 2006 in the UMTS network and was maintained in the LTE-SAE network and the second algorithm EIA2, based on AES algorithm, which will be studied in detail in section 3 of the present paper.

- Integrity key (IK) agreement;
- Data integrity feature: the receiving entity (ME or SN) must be able to check that the signaling data wasn't modified during its transition over the network access link and to check the expected origin of the message (SN (*Serving Network*) or UE).

## 1.3 EPS Algorithms Identification

As seen in the subsection 1.2.2 bellow, there are nowadays two sets of security algorithms used in the Long Term Evolution network. The first set is based on the stream cipher algorithm SNOW 3G, which is inherited for UMTS, the 3rd generation of mobile telecommunication. The second set is based on the well-known block cipher algorithm AES. Each algorithm is identified by an identifier. The following subsections present the EEA and the EIA algorithms identification used by 3GPP specifications documents [1].

### 1.3.1 EEA Algorithm Identification

The EPS Encryption Algorithms (EEA) are algorithms work with internal 128-bit blocks under the control of a 128-bit input key except Null ciphering algorithm.

To each EEA algorithm is assigned a 4-bit identifier. Currently, the following values have been defined for NAS, RRC and UP ciphering:

- "$0000_2$": EEA0 Null ciphering algorithm. The EEA0 algorithm is implemented in the way that it has the same effect as if it generates a keystream of all zeroes. The length of the generated keystream has to be equal to the LENGTH input parameter. Apart from this, all processing performed in association with ciphering are exactly the same as with any of the ciphering algorithms [2].

- "$0001_2$": 128-EEA1. The EEA1 is a stream cipher based on another stream cipher named SNOW 3G [6, 8, 9, 10]. As mentioned before, EEA1 is an inheritance from UTMS and was introduced as 3GPP standard on 2006.

- "$0010_2$": 128-EEA2. The EEA2 is a stream cipher based on the block cipher AES algorithm used in its CTR (*CounTeR mode*) mode.

UEs and eNBs shall implement EEA0, 128-EEA1 and 128-EEA2 for both RRC signaling ciphering and UP ciphering.

Besides, UEs and MMEs shall implement EEA0, 128-EEA1 and 128-EEA2 for NAS signaling ciphering.

It is important to note that the security functions are never deactivated, although it is possible to apply a NULL ciphering algorithm; The NULL algorithm may be used in certain special cases, such as for making an emergency call without a USIM.

### 1.3.2 EIA Algorithm Identification

Like EPS Confidentiality algorithms, all EPS Integrity Algorithms (EIA) works under control of a 128-bit input key, and for each one a 4-bit identifier is assigned. Currently, the following values have been defined [2]:

- "$0000_2$": EIA0 Null Integrity Protection algorithm. The EIA0 use is only allowed for unauthenticated emergency calls.
- "$0001_2$": 128-EIA1. The EIA1 is based on the stream cipher SNOW 3G [6, 8, 9, 10].
- "$0010_2$": 128-EIA2. The EIA2 is based on the block cipher AES used in its CMAC (Cipher-based MAC) mode.

The remaining values have been reserved for future use.

UEs and eNBs shall implement 128-EIA1 and 128-EIA2 for RRC signaling integrity protection. Both algorithms shall be implemented also by UEs and MMEs for NAS signaling integrity protection [2].

In the present paper, we are interested in studying the two algorithms EEA2 and EIA2 based on AES. Regarding the first set of cryptographic algorithms inherited from UMTS, their study has been subject of previous works [9, 10].

## 2. LTE Confidentiality algorithm EEA2

### 2.1 Encryption function EEA

The needs for a confidentiality protected mode of transmission are fulfilled by an LTE confidentiality cryptographic algorithm EEA [2, 5] which is a symmetric synchronous stream cipher. This type of ciphering has the advantage to generate the mask of data before even receiving the data to encrypt, which help to save time. Furthermore, it is based on bitwise operations which are carried out quickly.

Fig. 5 bellow illustrates the Encryption/Decryption operations using the algorithm EEA.

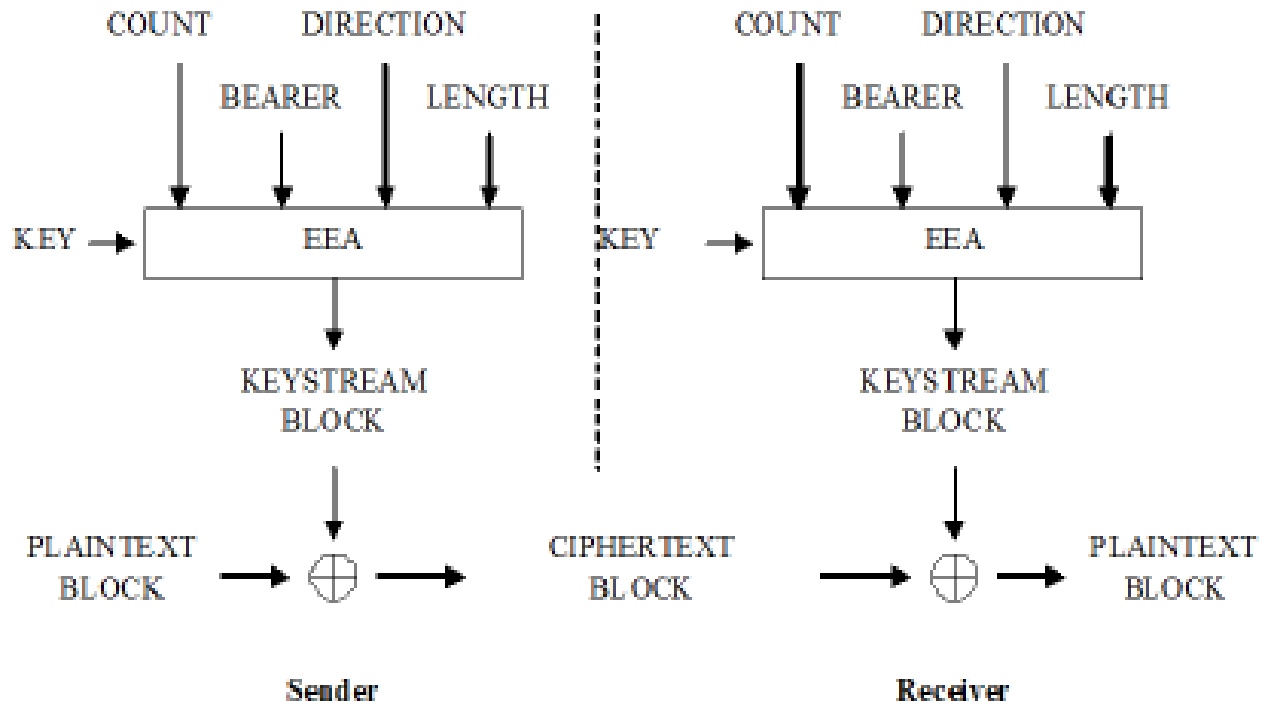

Fig. 5 Encryption/Decryption of user and signaling data

The input parameters of EEA are the same as for the UMTS encryption function f8. They are as follow:

- **COUNT-C**: Frame dependent input used to synchronize the sender and the receiver;
- BEARER : Service bearer identity;
- DIRECTION : Direction of the transmission;
- LENGTH : Number of bits to be encrypted /decrypted;
- Key: The cipher key. Unlike the UMTS where the encryption entities use only one cipher key for either user data or signaling data, in the EPS (LTE-SAE), different cipher keys are used depending on the data to protect:
  - $K_{UPenc}$: an 128-bit cipher key for User Plane confidentiality;
  - $K_{RRCenc}$: a 128-bit cipher key for RRC signaling data confidentiality, provided by the PDCP layer between UE and eNB (in the LTE network).
  - $K_{NASenc}$: a 128-bit cipher Key for NAS (Non-Access Stratum) signaling confidentiality, provided as part of the NAS protocol between UE and MME (in the SAE part).

Fig. 6 bellow shows the correspondence between security keys and information flows in the network [3].

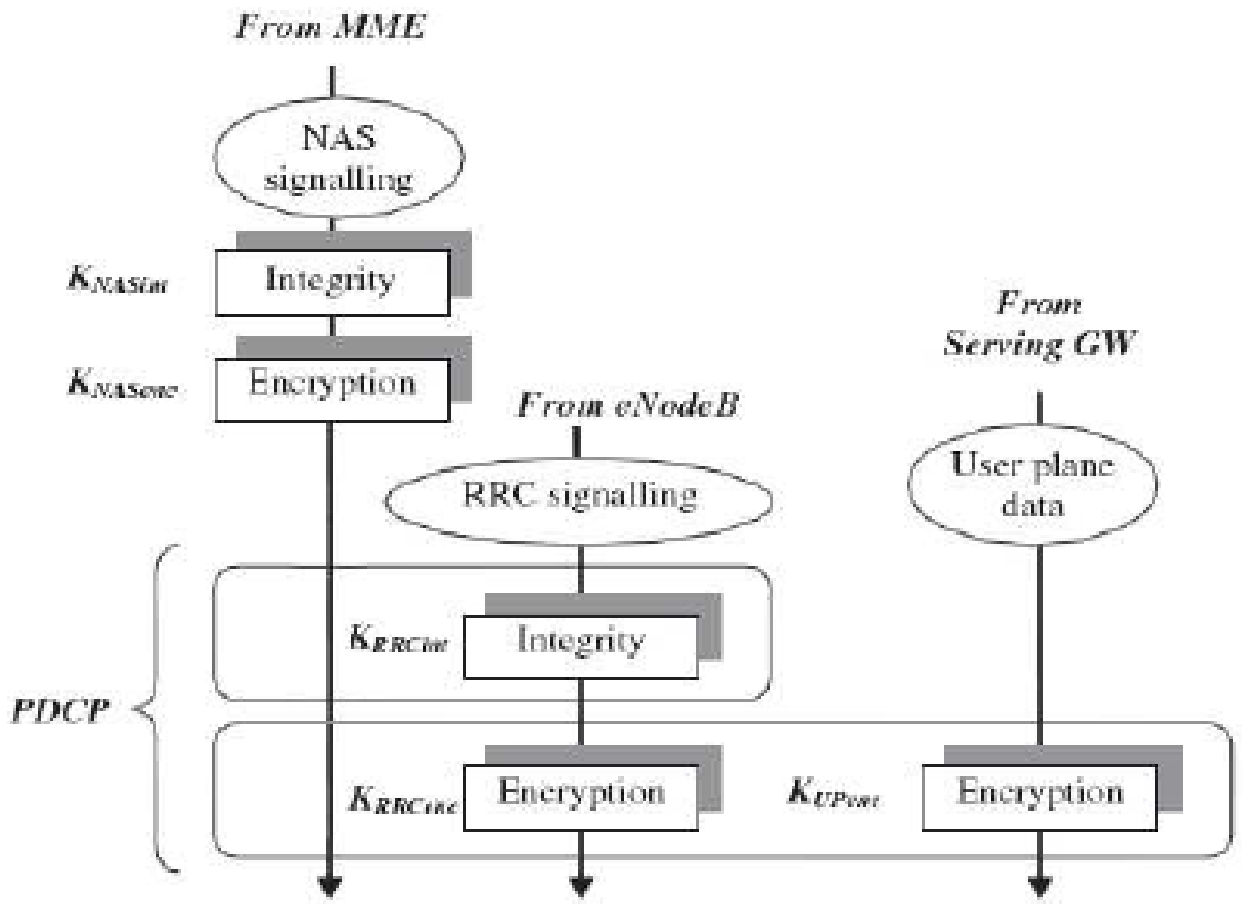

Fig. 6 Correspondence between security keys and information flows in the network [3]

For LTE, as seen above, two encryption algorithms EEA1 and EEA2 where adopted. The second variant EEA2 is based on the AES algorithm used in the CTR operation Mode (CTR Mode).

In the subsection bellow, we will be interested in studying the EEA2 operation and implementation.

## 2.2 Encryption Algorithm EEA2 Operation

The second EPS confidentiality algorithm EEA2 uses the block cipher AES as a kernel. The arguments for the choice of AES as a core algorithm for the second LTE Confidentiality algorithm (compared to UEA1 in the UMTS which uses the cipher block KASUMI) are given bellow. But apart from these, UEA1 based on KASUMI and EEA2 based on AES were perceived as equally good choices [1]:

- The eNB needs to support AES in any case because the eNB needs to support NDS/IP (Network Domain Security/Internet Protocol), which uses AES.
- The licensing conditions on the core of UEA1/UIA1 (Kasumi) do not make it free for use for other purposes than 3GPP access protection.
- Similarity with other non-3GPP accesses.

In this section, we will focus on the study of the EEA2 structure and operation to encrypt and decrypt messages. Indeed, 128-EEA2 is based on 128-bit AES [11] in CTR mode [12]. AES-CTR has many properties that make it an attractive encryption algorithm for high-speed networking. AES-CRT uses the AES block cipher to create a stream cipher. Data is encrypted and decrypted by using an exclusive-OR operation between intput data (the plaintext) and the keystream produced by encrypting sequential counter block values by the AES algorithm.

### 2.2.1 Initialization and keystream generation

The sequence of 128-bit counter blocks needed for CTR mode $T_1$, $T_2$, …, $T_i$, … is constructed as follows:

The most significant 64 bits of T1 consist of COUNT[0] .. COUNT[31] || BEARER[0] .. BEARER[4] || DIRECTION || $0^{26}$ (i.e. 26 zero bits). These input values are written from most significant bit on the left to least significant bit on the right, so for example COUNT[0] is the most significant bit of $T_1$. The least significant 64 bits of T1 are all 0. Fig. 7 bellow shows the counter construction from the EEA2 input data.

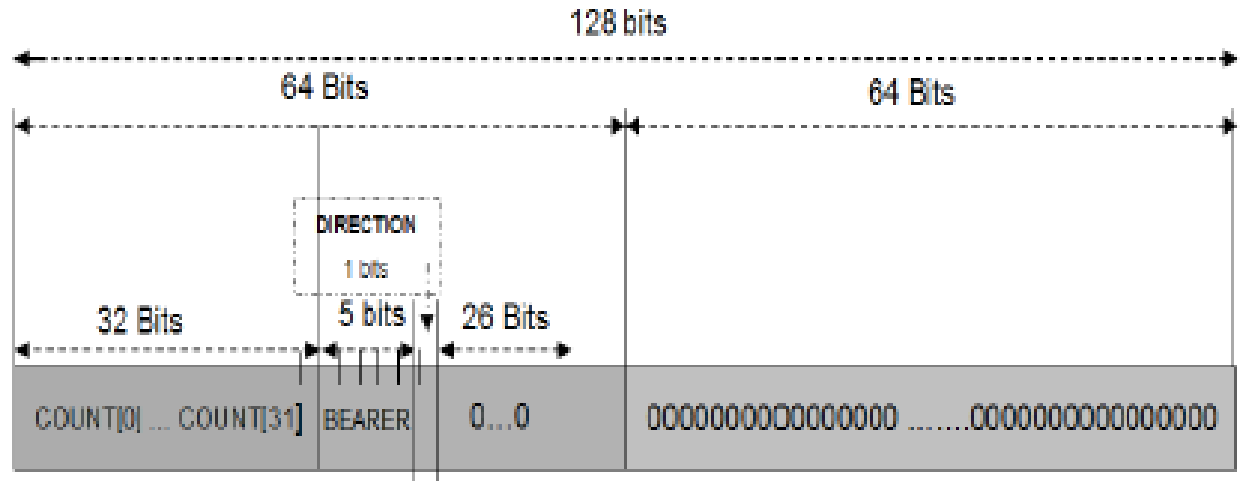

Fig. 7 First Counter block $T_1$

Subsequent counter blocks are then obtained by applying the standard integer incrementing function (according to Appendix B1 in [12]) mod $2^{64}$ to the least significant 64 bits of the previous counter block [2].

### 2.2.2 Encryption/Decryption

To encrypt a payload with AES-CTR, the encryptor partitions the plaintext, PT, into 128-bit blocks. The final block needs not to be 128 bits; it can be less.

PT = PT[1] PT[2] ... PT[n]

Each PT block is XORed with a block of the keystream to generate the ciphertext, CT. The AES encryption of each counter block results in 128 bits of keystream. The most significant 64 bits of the counter block T are initialized as seen before, followed by 64 bits that are all 0. This least significant 64 bits part of the counter T is the value that is incremented by one mod $2^{64}$ to generate subsequent counter blocks, each resulting in another 128 bits of keystream. The encryption of n plaintext blocks can be summarized as:

T := COUNT || BEARER || DIRECTION || $0^{26}$ || $T_0$ ($T_0$ indicate 64 bits all equal 0. It presents the part of the Counter block T that will be incremented mod $2^{64}$).

FOR i := 1 to n-1 DO

CT[i] := PT[i] XOR AES(T)

$T_0 := T_0 + 1$

END

CT[n] := PT[n] XOR TRUNC(AES(T))

The AES() function performs AES encryption under the control of the confidentiality key.

The TRUNC() function truncates the last output of the AES encrypt operation to the same length as the final plaintext block, returning the most significant bits.

Fig. 8 bellow illustrates the EEA encryption/decryption mechanism.

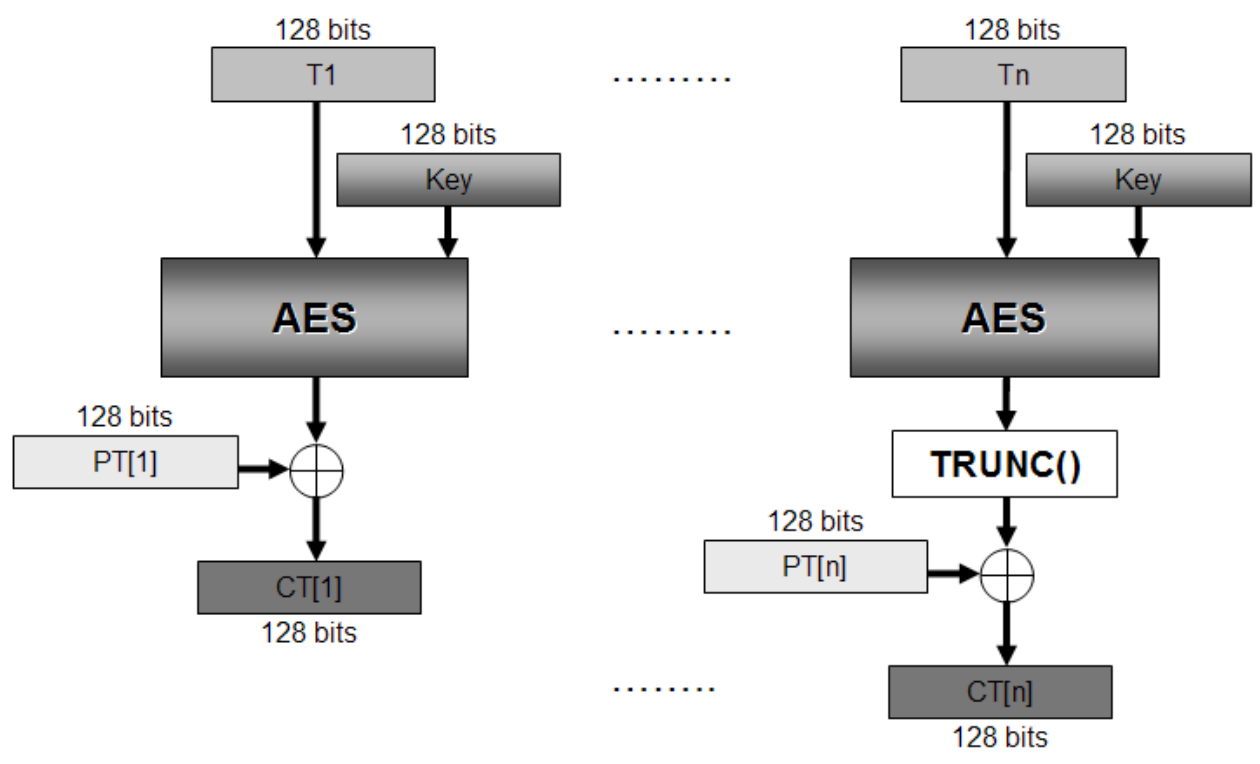

Fig. 8 EEA2 structure

The decryption operation is similar to the encryption and it is important to note that the AES-CTR uses the only AES encrypt operation (for both encryption and decryption), making AES-CTR implementation smaller than implementations of many other AES modes.

## 2.3 EEA2 Implementation

Unlike the first EPS algorithm EEA1 where the codes where given by the 3GPP specification, we have coded the EEA2 algorithm in C language with respect to endianess issues to avoid the memory reading problems faced with EEA1.

We have tested our implementation by performing all the TestSets given by 3GPP in the specification document [2], the results correspond to 3GPP expected results. We give you bellow one TestSet with the obtained result.

Input Data:

Key = 0a8b6bd8 d9b08b08 d64e32d1 817777fb
Count = 544d49cd  Bearer = 04
Direction = 0  Length = 310 bits

Plaintext =
fd40a41d 370a1f65 74509568 7d47ba1d 36d2349e
23f64439 2c8ea9c4 9d40c132 71aff264 d0f24800

Expected Ciphertext =
75750d37 b4bba2a4 dedb3423 5bd68c66 45acdaac
a48138a3 b0c471e2 a7041a57 6423d292 7287f000

Fig. 9 bellow shows the Test result which meets the 3GPP Test implementation data.

```
                EEA2 ALGORITHM ENCRYPTION

PROCESSING OF BLOCK1:

Counter block T1 = (hex) 544d49cd 20000000 00000000 00000000
Keystream block  = (hex) 8835a92a 83b1bdc1 aa8ba14b 2691367b
Plaintext block  = (hex) fd40a41d 370a1f65 74509568 7d47ba1d
Ciphertext block = (hex) 75750d37 b4bba2a4 dedb3423 5bd68c66


PROCESSING OF BLOCK2:

Counter block T2 = (hex) 544d49cd 20000000 00000000 00000001
Keystream block  = (hex) 737eee32 87777c9a 9c4ad826 3a44db65
Plaintext block  = (hex) 36d2349e 23f64439 2c8ea9c4 9d40c132
Ciphertext block = (hex) 45acdaac a48138a3 b0c471e2 a7041a57


PROCESSING OF BLOCK3:

Counter block T3 = (hex) 544d49cd 20000000 00000000 00000002
Keystream block  = (hex) 158c20f6 a275b8f5 0e8ae073 997c58ed
Plaintext block  = (hex) 71aff264 d0f248
Ciphertext block = (hex) 6423d292 7287f0
```

Fig. 9 EEA2 TestSet

# 3. LTE Integrity algorithm EIA2

## 3.1 Integrity function EIA

Integrity protection is realized by adding a field known as “Message Authentication Code for Integrity” (MAC-I) to each RRC or NAS message whose integrity has to be protected. This field is calculated by one of the EPS Integrity algorithms defined globally by the integrity function EIA. The input parameters to the integrity algorithm are the following:

- a 32-bit COUNT;

- a 5-bit bearer identity called BEARER;

- the 1-bit direction of the transmission i.e. DIRECTION. The DIRECTION bit shall be 0 for uplink and 1 for downlink;

- the message itself i.e. MESSAGE. The bit length of the MESSAGE is LENGTH;

- a 128-bit integrity key. As for the encryption key, in the EPS (LTE-SAE) system, different integrity keys are used, depending on the level where it is used. Fig. 6 presented in subsection 2.1 shows the different levels and keys used in each one.

  - $K_{RRCint}$: a 128-bit Integrity Key for the protection of the RRC signaling data integrity, provided by the PDCP layer between UE and eNB (in the LTE network).

  - $K_{NASint}$: a 128-bit Integrity Key for NAS (Non-Access Stratum) signaling integrity, provided as part of the NAS protocol between UE and MME (SAE network);

Fig. 10 illustrates the use of the integrity algorithm EIA to protect the messages integrity.

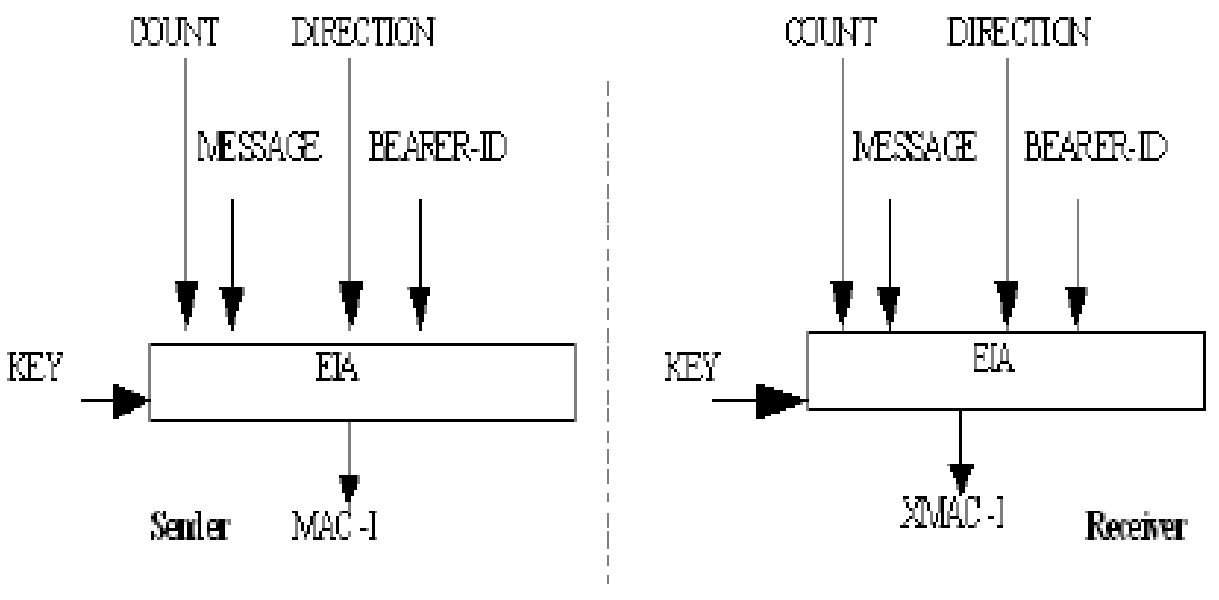

Fig. 10 Derivation of MAC-I/NAS-MAC (or XMAC-I/XNAS-MAC)

Based on these input parameters the sender computes a 32-bit message authentication code (MAC-I/NAS-MAC) using the integrity algorithm EIA. The message authentication code is then appended to the message when sent. The receiver computes the expected message authentication code (XMAC-I/XNAS-MAC) for the received message in the same way as the sender computed its message authentication code for the sent message and verifies the data integrity of the message by comparing it to the received message authentication code, i.e. MAC-I/NAS-MAC.

In the subsections bellow, we will be interested in the study of the EIA2 operation and then its implementation.

## 3.2 Integrity Algorithm EIA2 Operation

In this section, we will focus on the study of the EIA2 structure and operation to generate the MAC-I (*Message Authentication Code for Integrity*). Indeed, EIA2 is based on 128-bit AES [11] in the CMAC (cipher-based MAC) mode for Authentication [13]. CMAC, like any well-designed MAC algorithm, provides stronger assurance of data integrity than a checksum or an error detecting code. The verification of a checksum or an error detecting code is designed to detect only accidental modifications of the data, while CMAC is designed to detect intentional, unauthorized modifications of the data, as well as accidental modifications [13].

In the following subsections, we will present the different EIA2 operation steps.

### 3.2.1 Subkey Generation

Before proceeding to the MAC calculation, the EIA2 algorithm uses a subkey generation function to generate two subkeys needed during the MAC generation process.

Indeed, the integrity key K is used to derive two additional secret values, called the subkeys, denoted K1 and K2. The length of each subkey is 128 bits. The two subkeys are fixed once for any invocation of CMAC with the given key. So, for our implementation, the subkeys are precomputed and stored with the key for repeated use [13].

One of the elements of the subkey generation process is a bit string R which depends on the used block cipher AES size, which is in our case 128. So,

$$R_{128} = 0^{120}10000111$$

We present bellow, the specification of the subkey generation process of CMAC [13]:

$L = AES_K(0^{128})$
(128 is the block size of the block cipher AES used in EIA2)

IF $MSB_1(L) = 0$
    THEN $K1 = L << 1$;

ELSE
    $K1 = (L << 1) \oplus R_{128}$;

IF $MSB_1(K1) = 0$

THEN K2 = K1 << 1;

ELSE
K2 = (K1 << 1) $\oplus$ $R_{128}$

Return K1, K2.

### 3.2.2 Algorithm Initialization

As we have seen before, 128-EIA2 is based on 128-bit AES in CMAC mode. Before proceeding to the MAC-I calculation, the input parameters of the CMAC mode are constructed as follow:

The bit length of MESSAGE is LENGTH. The input to CMAC mode is a bit string M of length Mlen where Mlen = LENGTH + 64. M is constructed as follows ($M_Bit_i$ are the bits constituting M):

$M_Bit_0 \ldots M_Bit_{31}$ = COUNT[0] … COUNT[31]

$M_Bit_{32} \ldots M_Bit_{36}$ = BEARER[0] … BEARER[4]

$M_Bit_{37}$ = DIRECTION

$M_Bit_{38} \ldots M_Bit_{63} = 0^{26}$ (i.e. 26 zero bits)

$M_Bit_{64} \ldots M_Bit_{LENGTH+63}$ = MESSAGE[0] … MESSAGE[LENGTH-1]

### 3.2.3 MAC Generation

AES in CMAC mode is used with the inputs presented above to produce a Message Authentication Code T of length Tlen = 32. T is used directly as the 128-EIA2 output MACT[0] .. MACT[31], with MACT[0] is the most significant bit of T.

In the following, we present the generation process of the Message Authentication Code T, using the block cipher AES in the CMAC mode. We have exposed, in the subsection above, the input parameters of the algorithm. The bit representation of these parameters is $M_Bit_0 \ldots M_Bit_{LENGTH+63}$. However, it is important to note that during the MAC generation process, we will work with blocks of 128 bits constructed from the M_Bit bits sequence. These blocks are named M, such that $M = M_1 \| M_2 \| \ldots \| M_{n-1} \| M^*_n$, where $M_1, M_2, \ldots, M_{n-1}$ are complete blocks. These last remark is very usefully for understanding some steps of the MAC generation process, since some operations depends on the fact that the last block $M^*_n$ is a complete (containing 128 bits) or not complete (less than 128 bits) block.

Bellow is presented the MAC generation algorithm.

IF Mlen = 0
THEN n = 1;

ELSE
THEN n = $\lceil$Mlen/128$\rceil$;

IF $M^*_n$ is a complete block
THEN $M_n = K_1 \oplus M^*_n$;

ELSE
THEN {
j = n * 128 - Mlen - 1;
Mn = K2 $\oplus$ ($M^*_n \| 10^j$);
}

$C_0 = 0^{128}$;

FOR i = 1 to n
Ci = $AES_K (C_{i-1} \oplus M_i)$;

T = $MSB_{32} (C_n)$

RETURN T.

### 3.3 EIA2 Implementation

Like EEA2, the EIA2 algorithm code wasn't given by the 3GPP specification, so we have coded it in C language with respect to endianess issues to avoid the memory reading problems faced with EIA1.

We have tested our implementation by performing all the TestSets given by 3GPP in the specification document [2], the results correspond to 3GPP expected results. You find bellow an example of the implementation of one of the testsets given by 3GPP.

Input Data:

Key = 6832a65c ff447362 1ebdd4ba 26a921fe
Count = 36af6144 Bearer = 18
Direction = 0 Length = 383 bits

Message =
d3c53839 62682071 77656676 20323837 63624098
1ba6824c 1bfb1ab4 85472029 b71d808c e33e2cc3
c0b5fc1f 3de8a6dc

Fig. 11 bellow shows the Test result which meets the 3GPP Test implementation data.

```
      EIA2 ALGORITHM ENCRYPTION
           Based on AES-128
Mlen = 447
L  = e50123c387e13fd6 8d8bf0d0a4581685
K1 = ca0247870fc27fad 1b17e1a148b02d8d
K2 = 94048f0e1f84ff5a 362fc34291605b9d

MAC-I Generation:
n = 4
Mn* = c0b5fc1f3de8a6dc 0000000000000000
Mn  = 54b17311226c5987 362fc34291605b9d

C[0] = 0000000000000000 0000000000000000
M[1] = 36af6144c0000000 d3c5383962682071
C[1] = 263dd98fbeccb69a 428e92d421fbed9e
M[2] = 7765667620323837 636240981ba6824c
C[2] = 1838cb78cb2d32dc ec486c79d9007a19
M[3] = 1bfb1ab485472029 b71d808ce33e2cc3
C[3] = 5ebf1009f663be7b 683730724c20271f
M[4] = 54b17311226c5987 362fc34291605b9d
C[4] = f0668c1e4197300b 1243f83425d06c25

          MAC-I = f0668c1e
```

Fig. 11 EIA2 Implementation Example

## 4. Conclusion

In this paper, a detailed study of the EPS security mechanisms was carried out. We gave an algorithmic approach to this security subject. The objective was to understand and present clearly the 3GPP tendency regarding the cryptographic algorithms used to ensure data transmission security over 3G and 4G telecommunication networks. To achieve this and after our previous work on the first 3GPP set of cryptographic algorithms, we were interested, in the present paper, in the different modes and operation steps of the second set of LTE confidentiality and integrity algorithms EEA2 and EIA2 based on the block cipher AES.

We have presented the confidentiality algorithm EEA2 and the integrity algorithm EIA2 and then we have focused on their implementation in respect to the 3GPP requirements. Using the 3GPP implementation tests document, we tested our implementation and some results were exposed. We can realize that the algorithms EEA2 and EIA2 were well designed to be used for mobile telecommunication networks since the choice of their kernel algorithm (AES) and their operation modes ensures efficiency and rapidity, two criteria that are very crucial for such networks.

## References

[1] 3GPP specifications: http://www.3gpp.org

[2] 3GPP TS 33.401: “Technical Specification Group Services and System Aspects; 3GPP System Architecture Evolution (SAE): Security architecture; (Release 9)”, V9.1.0 (2009-09).

[3] P. Lescuyer, T Lucidarme, “Evolved Packet System (EPS) - The LTE and SAE Evolution of 3G UMTS”, John Wiley & Sons, Ltd, 2008.

[4] S. SECIA, I. Taoufik, M. BAKER, “LTE, The UMTS Long Term Evolution - From Theory to Practice”, John Wiley & Sons, Ltd. ISBN: 978-0-470-69716-0, 2009.

[5] 3GPP TS 33.102: "3rd Generation Partnership Project; Technical Specification Group Services and System Aspects; 3G Security; Security Architecture", December 2006.

[6] ETSI/SAGE Specification: “Specification of the 3GPP Confidentiality and Integrity Algorithms UEA2 & UIA2. Document 1: UEA2 and UIA2 Specification”. Version: 1.1, September 2006.

[7] I. Pooters, University of Twente, Faculty of EEMCS: “An Approach to full User Data Integrity Protection in UMTS Access Networks”, April 2006.

[8] ETSI/SAGE Specification: “Specification of the 3GPP Confidentiality and Integrity Algorithms UEA2 & UIA2. Document 2: SNOW 3G Specification”, Version: 1.1, September 2006.

[9] G. Orhanou, S. El Hajji, Y. Bentaleb, "SNOW 3G Stream cipher Operation and Complexity Study", Contemporary Engineering Sciences - Hikari Ltd, Vol. 3, 2010, no. 3, 97 - 111, 2010

[10] G. Orhanou, S. El Hajji: "UTRAN Cryptrographic algorithms - Operation and complexity study", Journal of Theoretical and Applied Information Technology (JATIT), Slot number 4, April 2010, Volume 14 n°2, 97-106. 2010

[11] FIPS Publication 197, “Advanced Encryption Standard (AES)”, U.S. DoC/NIST, November 26, 2001.

[12] NIST Special Publication 800-38A, “Recommendation for Block Cipher Modes of Operation - Methods and Techniques”, December 2001.

[13] Morris Dworkin, NIST Special Publication 800-38B, Recommendation for Block Cipher Modes of Operation: The CMAC Mode for Authentication May 2005.